# UNDEREXPANDED SUPERSONIC JET IN IMPOSED OSCILLATING CONDITION

Md. Elius, Md. Mahmudul Hasan, and A. B. M. Toufiqe Hasan[*]

Department of Mechanical Engineering, Bangladesh University of Engineering and Technology (BUET)
Dhaka-1000, Bangladesh
*E-mail of corresponding: toufiquehasan@me.buet.ac.bd

**Abstract**

In the present study, a computation study is performed to investigate the effect of imposed oscillation of nozzle pressure ratio (NPR) on the flow structure in a two-dimensional, axisymmetric supersonic converging nozzle. In this study, the underexpanded flow conditions are considered which are dominated by diamond shock-cell structure. The computational results are well validated with the available experimental measurements. The flow is initially computed to be fully developed and then oscillations are imposed. NPR is increased from 1.6 to 2.6 and then decreased again to 1.6 and thus completes a cycle. Results showed that the external flow structure of the nozzle is dependent on the process of change of pressure ratio during the oscillation. Distinct flow structures are observed during increasing and decreasing processes of the change of pressure ratio even when the nozzle is at the same NPR. Irreversible behaviors in the locations of jet centreline axis and off-axis as well as expansion, compression and neutral zones, are observed at the same NPRs during this oscillation. Further, the effect of oscillation frequency is explored on this irreversible behavior at 100 Hz, 200 Hz, 500 Hz and 1000 Hz frequencies.

## 1. Introduction

The research on flow through a supersonic converging nozzle is of great importance in many advanced engineering applications, especially in the aeronautical and aerospace industries. Under-expanded jets are complex high speed flows, which are formed in various engineering applications and devices such as exhaust plumes of aircrafts (rockets and missiles) [1], supersonic combustors, actuators, ejectors high pressure gaseous injectors and soot blower jet impinging [2] and explosive volcanic eruptions [3]. An under-expansion may occur whenever a fluid is released from a device at a pressure greater than the ambient pressure. Much work has been dedicated to it, particularly the ratio of the nozzle exit pressure to the ambient pressure, which influences the jet structure. It is an important factor used to characterize supersonic jets.

Franquet et al. [4] on their free underexpanded jets review performed an exhaustive overview of the main experimental papers dealing with underexpanded jets, this paper aim was to clarify the characteristics which are well known from those where is clearly a lack of confidence. In many real field applications such as airblast nozzle(injector) the pulsating conditions are observed. Strasser [5] investigated the characteristics of three-stream co-axial airblust nozzle in the experiments and computations with frequencies between 75 Hz and 600 Hz which is used to generate an atomized fuel stream for a large-scale reactor. On the basis of frequencies, three distinct flow patterns was observed. By continuously changing the back pressure with time, Irie et al. [6] focused on characteristics of the Mach disk in the underexpanded jet. Furthermore, it is known that the hysteresis phenomena of a Mach disk is affected by the pressure ratio [7]. Many papers have described the hysteresis phenomena for an under-expanded jet [8]. Various attempts have been made to know the flow behavior of under-expanded jets.

The purpose of this study is to investigate the hysteresis phenomena of shock-cell structure in under-expanded axisymmetric supersonic jet during the transient processes of change in the nozzle pressure ratio.

## 2. Numerical Method and Validation

In this paper, a two-dimensional model (2D) axisymmetric model is considered instead of three-dimensional (3D) model for quick modelling and efficient analysis. Qin et al. [9] presented in their paper that plane axisymmetric and 2D planer models can provide efficient, reliable, and accurate results as compared to the analytical results of equivalent 3D models. The flow material is air and in the present study air is considered as an ideal gas. The flow is considered to be steady and viscous. The flow is governed by the Navier-Stokes equations. The turbulence model utilized is $k$-$\omega$ SST, which require two additional transport equations to be solved. Finite volume method is used to spatially discretize the equations for numerical simulation. Fig. 2a shows the model geometry. The throat diameter ($D$) is 38.25 mm and the compression area ratio is 2.6. Fig. 2b shows the computational domain and boundary conditions. The origin of ($x$, $y$) coordinate system is located at the center of the throat. The nozzle pressure ratio (NPR) $p_{01}/p_b$, where $p_{01}$ is inlet total pressure and $p_b$ is the back pressure, is oscillated in triangular wave as shown Fig. 1. The back pressure is kept constant at 100 kPa. The oscillation of NPR consists of two processes namely-increasing process. (NPR = 1.6 – 2.6) and decreasing process (NPR = 2.6 – 1.6). Four cases are considered by varying the frequency of oscillation. These are: $f$ = 100 Hz, 200 Hz, 500 Hz and 1000 Hz. The corresponding time periods are 10 ms, 5ms, 2 ms and 1 ms, respectively. The inlet total temperature is kept at 300 K. Considering the computational time, a grid size of 250,175 was found optimum and used for all the computations in the present study. In this case, the minimum normal grid spacing was reduced to give the value of wall $y+ \approx 1.0$. A time step size of $10^{-5}$ s was found sufficient for this type of unsteady computation. Moreover, 500 internal iterations are performed at each time step to ensure the convergence of the solution variable.

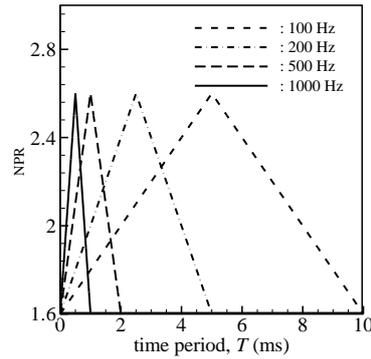

Fig. 1: Variation of NPR with time

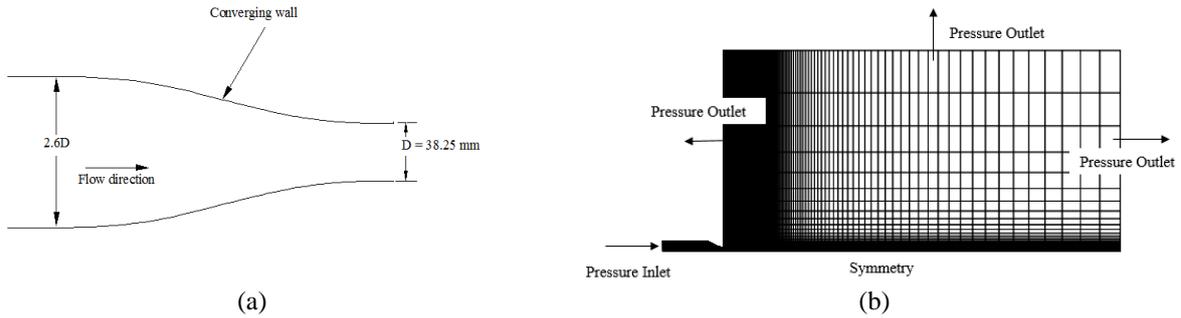

(a)           (b)

Fig. 2: (a) Nozzle geometry (b) Computational domain with boundary conditions.

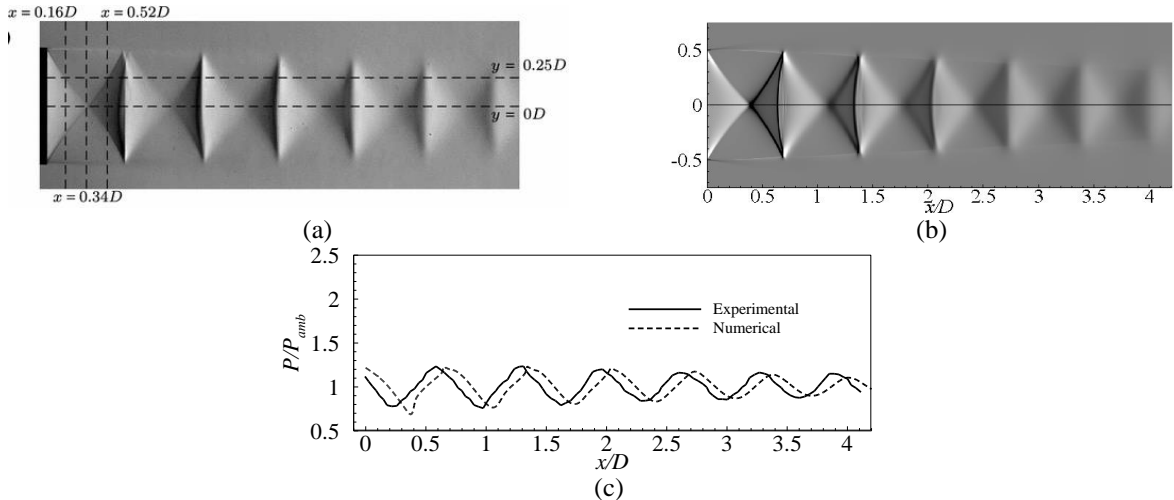

Fig. 3: Comparison of the Schlieren images of shock-cell structure for both (a) experiment [10] and (b) present numerical computation and (c) the static pressure at jet centerline

The performance of the present computational method is verified against available experimental results obtained by André et al. [10]. In the experiment, the inlet pressure was 2.27 bar and outlet pressure was 1 bar. In the following sections the Schlieren image of the shock-cell structure and pressure distribution at the center line are compared with experimental data. In Fig. 3 comparison of Schlieren images is made between the experimental (Fig. 3(a)) and present numerical computation (Fig. 3(b)). The number of shock-cell is same for both experimental and present numerical computation. In every shock-cell both expansion and compression region slightly move downstream in present numerical computation as confirmed from pressure distribution (Fig. 3(c)). Accordingly, it can be concluded that the present computational methods works well for such underexpanded nozzle flow conditions.

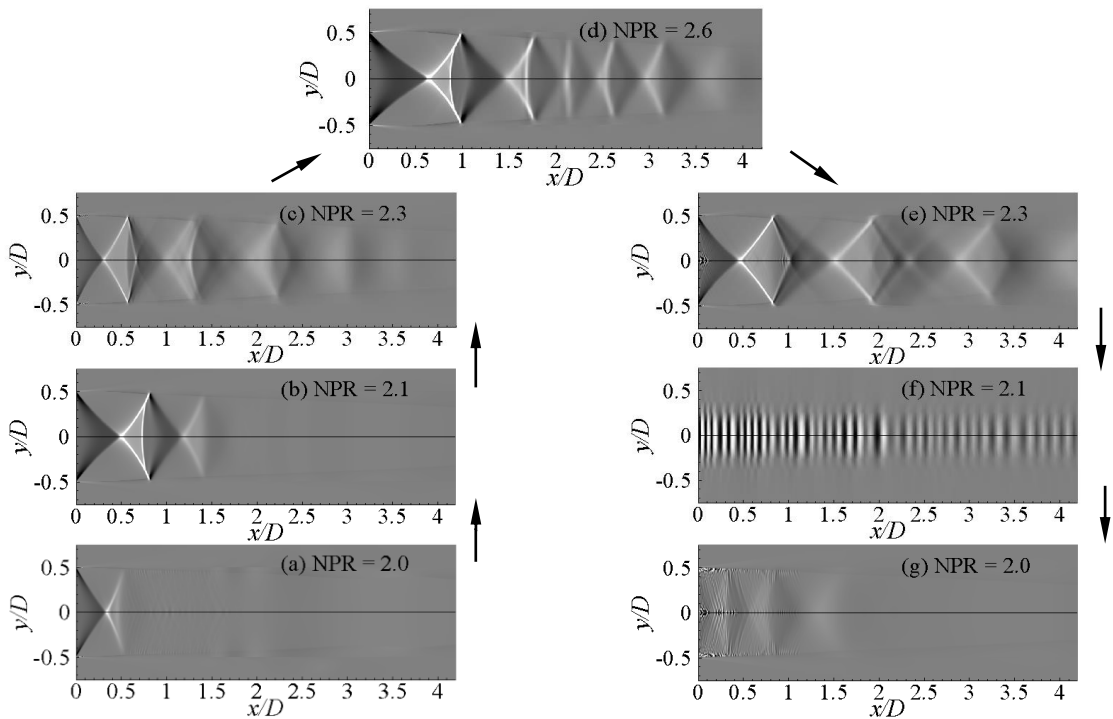

Fig. 4: Schlieren images at different NPR at 100Hz

In case of oscillating conditions, the computed results are taken from $3^{rd}$ cycle of NPR oscillation for $f$ = 100Hz and $f$ = 200Hz. However, $5^{th}$ cycle has been considered for $f$ = 500Hz and $f$ = 1000Hz since the results were different in initial cycles.

## 3. Results and Discussion

Fig. 4 shows schlieren images of flow field during the successive increasing and decreasing modes of change of pressure ratio for oscillation frequency, $f$ = 100 Hz. The left side sequence presents the increasing process of NPR and the right side for the decreasing process of pressure ratio. At the beginning of increasing process of NPR = 2.0 (Fig. 4a), the expansion wave of first shock cell is located at $x/D \approx 0.35$. With the successive increasing of NPR, at NPR = 2.1(Fig. 4b) the expansion wave moves downstream but NPR = 2.3 (Fig. 4c) shows exceptional. However, at higher NPRs of 2.3 and 2.6, the shock cells are well developed. At the end of increasing process of NPR = 2.6 (Fig. 4d), the expansion wave of first shock cell is located at $x/D \approx 0.6$. After that the NPR starts to decrease. However, the response of pressure decrease on the flow structure is not spontaneous. At NPR = 2.3 (Fig. 4e) the expansion wave of first shock cell is located at $x/D \approx 0.5$ which is quite downstream to the corresponding of location of $x/D \approx 0.25$ in increasing process (Fig. 4c). Then the flow structure emerges without shock cell at NPR = 2.1 (Fig. 4f) and the end of the decreasing process of NPR 2.0 (Fig. 4g), the flow structure emerges with slightly visible two shock cell which is different from in increasing process of NPR = 2.0. As seen from these picture sequences, there are differences between the locations of first shock wave even in the same pressure ratio.

Results of schlieren pictures of for the case of $f$ = 200 Hz are shown in Fig. 5. In this case, at the beginning of increasing process at NPR = 2.0 (Fig. 5a), shock cell isn't present here which is completely different compared to the case of $f$ = 100 Hz (Fig. 4a). However, at next NPR of 2.1 (Fig. 4b) in the successive process, the flow field is stronger without shock cell from NPR = 2.0 (Fig. 5a). Thus it can be seen at higher oscillating condition, the response of initial pressure ratio increment on flow structure is seen to have a lag compared to the lower frequencies (for example Fig. 4). After that with the successive increasing of NPR, at NPR = 2.3 (Fig. 5c) one shock cell come out. At NPR = 2.6 (Fig. 5d) the expansion wave of first shock cell is located at $x/D \approx 0.75$. Thus ends the increasing process. Then the NPR starts to decrease at the initial NPR = 2.3 (Fig. 5e) the flow structure is different in case of shock cells position compared to the increasing process of NPR = 2.3 (Fig. 5c). With further decrease in pressure ratio NPR 2.1, 2.0 (Fig. 5f-g) the flow structure is emerged with strong shock cells which are completely distinct from increasing process of NPR 2.1, 2.0 (Fig. 5b-a).

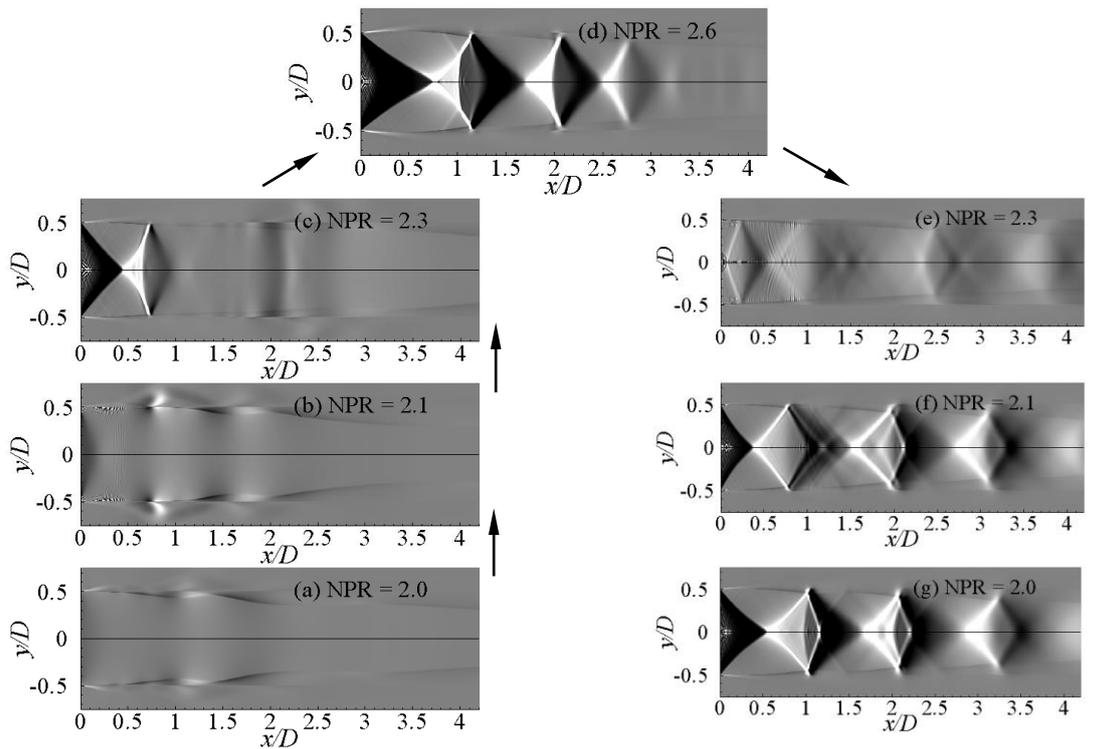

Fig. 5: Schlieren images at different NPR at $f$ = 200Hz

The flow structure for the case of $f$ = 500 Hz and 1000 Hz also correspond to the similar irreversible behavior during the oscillation of NPR. These sequential figures are not shown here for brevity. However, the radial profile of axial velocities will be shown in the following discussion.

Now, the static pressure and axial velocity variation at $f$ = 100 Hz at jet centerline will be discussed. Fig. 6 and Fig. 7 show the static pressure and axial velocity variation at jet centerline during the oscillation of NPR. Solid line presents the increasing process and the long dash presents the decreasing process. NPR =2.0, 2.1, 2.3 (Fig. 6a-c and Fig. 7a-c) are presented. To describe the static pressure and axial velocity variation (for example increasing process of NPR 2.3), expansion occurs within the light- pointing triangles (Fig. 4c) of the schlieren image: the static pressure falls and the axial velocity increases. Conversely, compression takes place in the dark left-pointing triangles (Fig. 4c) since static pressure increases and axial velocity falls. This pattern is repeated until the end of the potential core (not visible here). The static pressure curve is seen to oscillate about the ambient pressure in a slightly damped fashion. However, the decreasing process curve of NPR = 2.3 (Fig. 6c) doesn't follow the increasing curve; shows the hysteresis. Fig. 6 clearly shows that with successive increasing of NPR, the static pressure fluctuation and hysteresis increases as seen in Fig. 6a-c for NPR = 2.0, 2.1 and 2.3 respectively. The axial velocity at jet centerline also shows the similar irreversible behavior during the oscillation of NPR. The flow structure for the case of $f$ = 200 Hz, 500Hz and 1000Hz also correspond to the similar irreversible behavior during the oscillation of NPR. These sequential figures are not shown here for brevity.

Fig. 8 shows variation of axial velocity in radial direction at $y/D$ = .16D for the cases of $f$ = 100 Hz (Fig. 8a), 200 Hz (Fig. 8b), 500 Hz (Fig. 8c) and 1000 Hz (Fig. 8d) at NPR = 2.0. At the increasing process of $f$ = 100 Hz (Fig. 4a), where the centerline velocity increases, a velocity dip at the centerline can be noticed. This due to the fact that the expansion fans are attached to the lip so that their influence reaches the centerline further downstream. This means that the fluid acceleration occurs sooner in the outer part of the jet as compared to the centerline. These behavior clear shows that $y/D$ = 0.16D located at expansion zone. However, in decreasing process (Fig. 8a), the velocity is nearly constant except the jet centerline. At $f = 200$ Hz (Fig. 8b), the velocity is nearly constant in increasing process but in decreasing process the centerline velocity catches up with outer velocity and finally moves past it compared to the increasing process curve of $f = 100$ Hz (Fig. 8a) and it is nearly located at neutral zone according these behaviors. Both in increasing and decreasing process of $f = 500$ Hz the axial velocity distribution becomes almost same. Both are located at expansion zone similar to the increasing process of $f = 100$ Hz (Fig. 8a). At the last cases $f = 1000$ Hz (Fig. 8d) in increasing process, the velocity variation is higher compared to the increasing process of $f = 100$ Hz and $f = 500$ Hz. But, the velocity variation in decreasing process is nearly constant. Adding all the successive locations, a hysteresis can be found for all the flow cases as shown in Fig. 8. The intensity of hysteresis is seen to increase at higher oscillation frequencies except $f = 500$ Hz. The axial velocity in radial direction at $y/D = 0.16D$ for the NPR = 2.1, 2.3 and 2.6 also correspond to similar irreversible behavior. These sequential figures are not shown here for brevity.

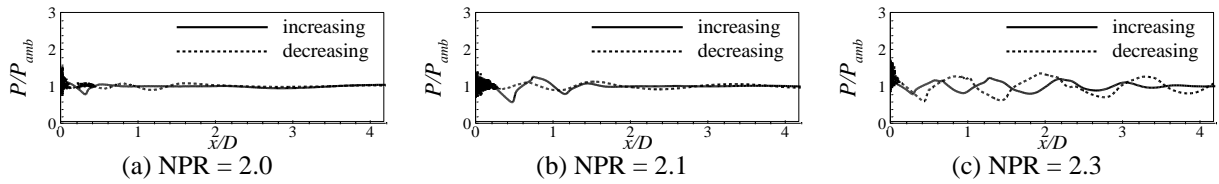

Fig. 6: Static pressure variation at jet centerline at 100Hz

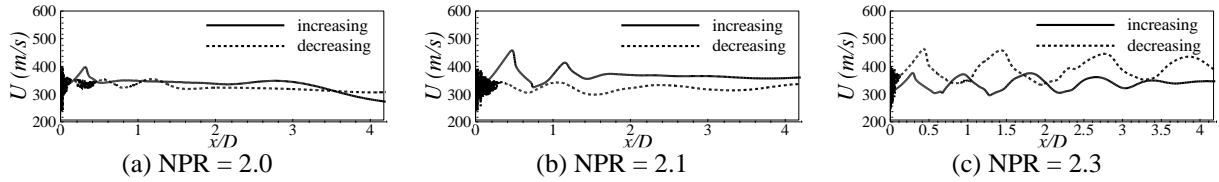

Fig. 7: Axial velocity variation at jet centerline at $f = 100$ Hz

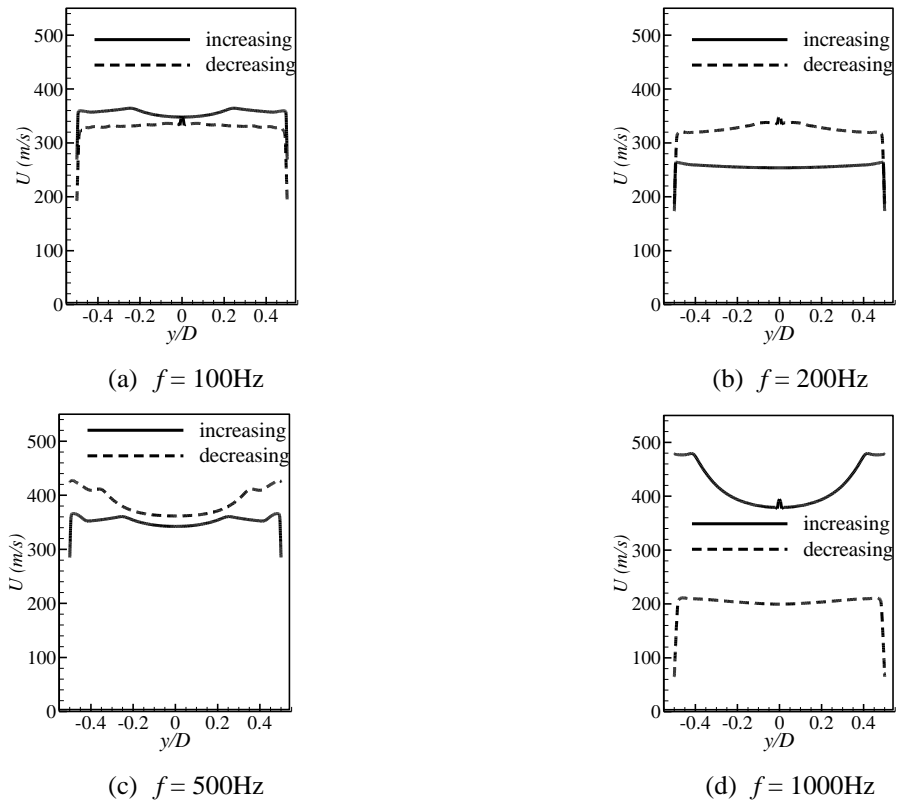

Fig. 8: Axial velocity distribution along radial direction at $y/D = 0.16D$

## 4. Conclusions

A numerical simulation is performed to investigate the effect of the oscillation of nozzle pressure ratio (NPR) on the shock-cell structure outside a two-dimensional, axisymmetric converging supersonic nozzle. Underexpanded supersonic jet conditions are considered which is dominated by the shock-cell structure. The effect of oscillation frequency is also explored in this study. Computational results are well validated with the available experimental data. The results of the present study can be summarized as follows:

(i) The flow structure exhibits an irreversible characteristics during the NPR oscillation. The locations of jet centreline axis and off-axis as well as expansion, compression and neutral zones are significantly different in increasing and decreasing processes even at the same NPR during the oscillation of nozzle pressure ratio.
(ii) The successive locations of the shock structure during increasing and decreasing processes produces a hysteresis loop in a complete cycle of NPR oscillation
(iii) The computed results shows that the hysteretic characteristics are increased at higher oscillation frequencies.